\documentclass{article}
\usepackage{amssymb}
\usepackage{bm}
\usepackage[dvips]{graphicx}
\begin{document}

\title{Efficiency of spin-injection terahertz oscillator}
\author{Yu.V. Gulyaev$^1$, S.G. Chigarev$^1$,
I.V. Malikov$^2$,\\ G.M. Mikhailov$^2$, P.E. Zilberman$^1$, E.M. Epshtein$^1$\thanks{E-mail: epshtein36@mail.ru}\\ \\
$^1$\emph{V.A. Kotelnikov Institute of Radio Engineering and Electronics}\\
    \emph{of the Russian Academy of Sciences, 141190 Fryazino, Russia}\\ \\
    $^2$\emph{Institute of Microelectronics Technology and High Purity Materials} \\
    \emph{of the Russian Academy of Sciences, 142432 Chernogolovka, Russia}}
\date{}
\maketitle

\abstract{Energy efficiency in terahertz range is evaluated experimentally of a
spin-injection oscillator based on a ferromagnetic rod-film structure with
point contact between the components. Choice of the film material
influences substantially the efficiency. A magnetic flux concentrator is
used to improve the efficiency. It is found from the measurements that the
quantum efficiency can exceed unity. The latter indicates substantial
contribution of stimulated radiative transitions.} \\ \\

\section{Introduction}\label{section1}
Investigations and applications of terahertz (THz) radiation evoke increasing
interest. Such a radiation interacts softly with biological objects
without radiative damage, it is suitable for nondestructive diagnostics,
it spans actual range of a wide class of internal particle motions in
materials, that makes it possible to study material structure and
properties, to miniaturize sizes and weight of the local communication
devices, etc.~\cite{Wang}. Broad use of THz devices is hindered nowadays
because of absence of compact and reliable sources. A frequency converting method is used
that is cumbersome and low-effective~\cite{Li,Cook}.

Recently, an idea has been proposed of a laser-type coherent THz oscillator
based on spin-polarized current through a magnetic
junction~\cite{Kadigrobov1,Kadigrobov2}. However, some problems of
principal and technological character appear in attempts of realizing the
idea~\cite{Gulyaev1,Gulyaev2}. To overcome the problems, new types of
layered structures were proposed with a point contact between conducting
ferromagnetic rod and very thin ferromagnetic metal film. Such structures
were realized and tested. Calculations show possibility of reaching very
high current density $\sim10^7$--$10^9$ A/cm$^2$ without damage and
overheating~\cite{Gulyaev3}. The experiments confirmed these
estimates~\cite{Gulyaev3,Gulyaev4}. Besides, a possibility was predicted
theoretically of reaching negative effective spin temperature and
generation of THz radiation~\cite{Gulyaev4}. Recently, spin-injection
driven THz radiation has been observed~\cite{Gulyaev5}.

In this work, the results are presented of experimental study of integral
energy characteristics of the radiator and the construction described
in~\cite{Gulyaev5}. Also, attention is paid to estimates of the radiator
quantum efficiency based on the measurements.

\section{The experimental set}\label{section2}

\begin{figure}
\includegraphics[width=100mm]{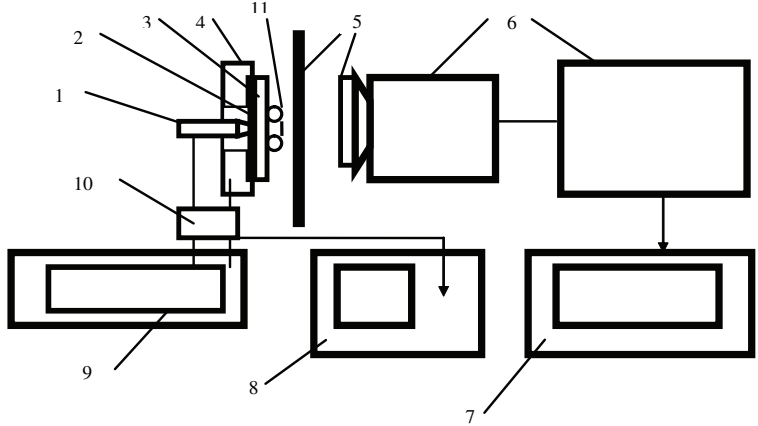}
\caption{Scheme of the experimental set. The block numbers are explained
in the text.}\label{fig1}
\end{figure}

The measurements were carried out with experimental set shown in
Fig.~\ref{fig1}. The radiator consists of a sharpened steel rod 1,
ferromagnetic film 2 deposited onto a dielectric substrate 3 transparent
in THz range, and an electrode 4 closing the electric circuit. The signal
passed through the substrate was recorded with power meter based on a
Golay cell 6. The THz range is determined by means of two filters 5,
namely, a low frequency filter of a metal grid with $125\times125$
$\mu$m$^2$ meshes and a TYDEX polymer high frequency filter. The radiator
operating parameters (current and voltage) were recorded by a digital
oscillograph 8, and the radiation power by a digital recorder. A dc source
was used which supplied operating current up to 1 A. A resistance block 10
protected the power source 9 against overload under circuit shorting and
formed signals in measurements. In some experiments, a magnetic field
concentrator 11 was used.

\section{The role of material}\label{section3}
\begin{figure}
\includegraphics[width=100mm]{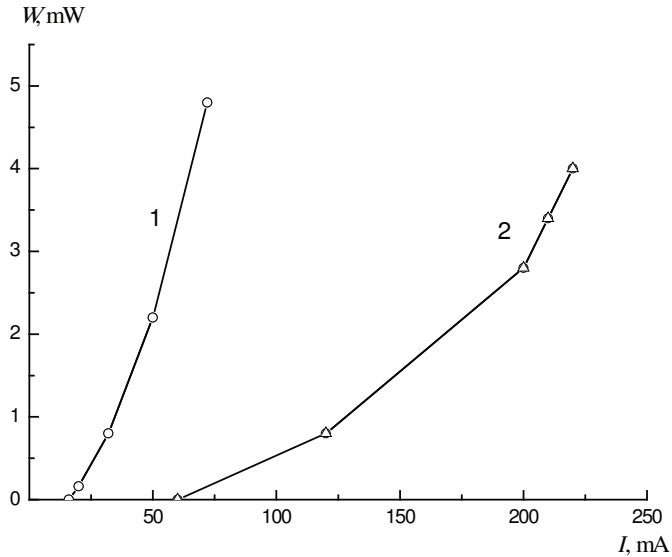}
\caption{The radiation power as a function of the current: 1~---
with Fe$_3$O$_4$ film, 2~--- with Permalloy film.
}\label{fig2}
\end{figure}

Epitaxial ferromagnetic films of two types were grown on the r plane of single
crystalline sapphire by pulsed laser
evaporation, namely, polycrystal Permalloy (Py) films 30 nm thick and
epitaxial magnetite (Fe$_3$O$_4$) films 250 nm thick. Epitaxial Fe$_3$O$_4$
film was grown at $340^\circ$ C on a sapphire (Al$_2$O$_3$(-1012)) with
MgO(001) underlayer 10 nm thick by laser evaporation of a Fe target in
molecular oxygen at $9\times10^{-5}$ Torr pressure. The radiator with the films
indicated was studied. The radiation power as a function of the current is
shown in Fig.~\ref{fig2}. The curves 1 and 2 correspond to radiator
with Fe$_3$O$_4$ and Py films, respectively. The maximal values of the
currents in the experiments were restricted with breakdown in the film to
rod contact. It is seen that the maximal radiation power was about 5 mW
for both curves. However, to reach such a power, half as large current is
required with Fe$_3$O$_4$ film, as with Py film. It is interesting to
compare corresponding starting current values needed to observe radiation.
It appears that the starting current is three times less with using
magnetite film in comparison with Py film. It shows that magnetite allows
more effective operation of the radiator than Py. We consider this with
more detail below.

\section{Estimates of efficiency}\label{section4}
\begin{figure}
\includegraphics[width=100mm]{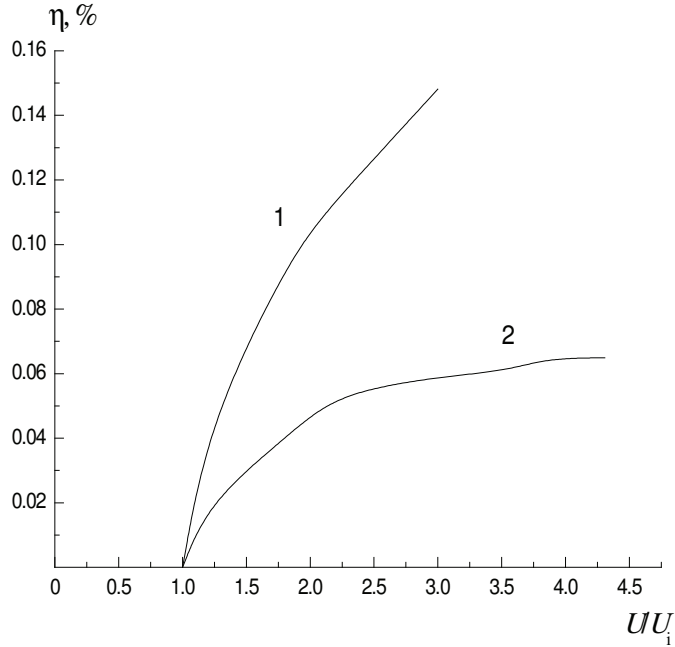}
\caption{Energy efficiency as a function of the ratio of $U$ voltage drop in
contact to the starting voltage $U_i$: 1~--- with Fe$_3$O$_4$ film, 2~--- with Permalloy film.
}\label{fig3}
\end{figure}

Measurements of the voltage drop $U$ in the magnetic junction and current
$I$ allow to evaluate the power released by the current $W_c=UI$ and the
energy efficiency of the radiator $\eta\equiv W/W_c$, where $W$ is the
radiation power. Such efficiencies for two film types are shown in
Fig.~\ref{fig3}. The maximal efficiency reaches 0.15\% with the magnetite
film, and 0.06\% with the Py film. The difference is due to different
equilibrium spin polarization $P$ that is about 0.4 for Py and is near to
1 for Fe$_3$O$_4$~\cite{Tripathy}. Also, it should have in mind that
magnetite is closer to semiconductor compounds, in its properties, than
Py, so that larger spin diffusion length may be expected for
magnetite~\cite{Tripathy}. Therefore, the interaction length of active
spins with radiation may be larger, too, that leads to higher efficiency~\cite{Gulyaev5}.

\section{Influence of concentrator}\label{section5}
\begin{figure}
\includegraphics[width=100mm]{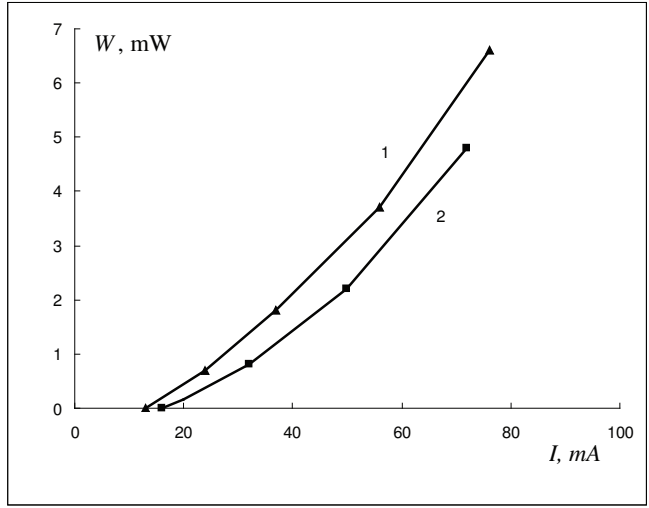}
\caption{The concentrator effect on the radiating power in the structure
with Fe$_3$O$_4$ film: 1~--- with concentrator ($D=1.5$ mm, $d=1$ mm),
2~--- without concentrator.}\label{fig4}
\end{figure}

There is a possibility to control the radiation power and energy
efficiency by varying closing magnetic flux in the
radiator~\cite{Gulyaev5}. The flux depends substantially on the magnetic
core 11 under the substrate 3. The magnetic flux may be carried away from
the sample completely with a layer of soft magnetic (transformer) steel. As a result, the
spin-injection driven radiation disappears. It is possible, also, to
enhance the radiation with such a magnetic core by concentrating the
closing magnetic flux on the sample. For example, we made a magnetic concentrator
in the form of a ring with diameter $D=1.5$ mm of an iron wire with
diameter $d=1$ mm. The dependence of the radiation power on the current
with using this concentrator is shown in Fig.~\ref{fig4}. It is seen, that
the concentrator decreases the starting current and leads to increasing
the power under the same current value. The presented data are not optimal
ones, they merely illustrate possible influence of concentrator.

\begin{figure}
\includegraphics[width=100mm]{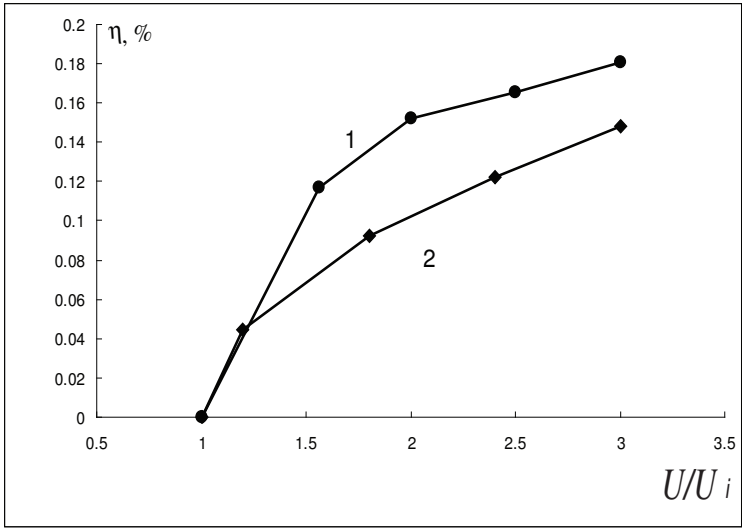}
\caption{The energy efficiency of the radiator as a function of relative
voltage drop in the contact $U/U_i$ in the structure
with Fe$_3$O$_4$ film: 1~--- with concentrator ($D=1.5$ mm, $d=1$ mm),
2~--- without concentrator.}\label{fig5}
\end{figure}

The energy efficiency of the radiator with Fe$_3$O$_4$ film as a function
of the ratio of the operating voltage to the starting one with and without concentrator is shown in
Fig.~\ref{fig5}. All the results indicate that the magnetic flux
concentrator can enhance efficiency of forming negative spin polarization
in the magnetic junction and increase the radiation power.

\section{Quantum efficiency}\label{section5}
Here, the quantum efficiency $\theta$ means the number of radiated photons per a
current carrier. This quantity is interesting because it shows ``radiative
capability'' of the relaxation processes in the device and characterizes
intensity of radiative processes. It can be calculate from experimental
data as a ratio
\begin{equation}\label{1}
  \theta=\frac{W}{\left|I/e\right|\hbar\omega},
\end{equation}
where $e$ is the electron charge, $\omega$ is the radiation angular
frequency, $\hbar$ is the Planck constant. By substituting to Eq.~(\ref{1})
the data for Fe$_3$O$_4$ from Fig.~\ref{fig3}, namely, $W=4.8$ mW, $I=72$
mA, $\omega=2\pi\times10^{13}$ s$^{-1}$, we obtain $\theta\approx1.7$.

Thus, the quantum efficiency appears to be more than unity. For the Py
film, this parameter is lower, but may be close to unity. It should have
in mind that non-radiative processes take place at room temperature. So,
it may be concluded that a regular cause exists of the enhancement of the
quantum efficiency. In our opinion, such enhancement may be due to substantial
contribution of stimulated processes, which are induced by the
electromagnetic energy stored in the radiator. The energy accumulation is
a result of reflection the waves radiated earlier from the interfaces. It
may be expected that placing the radiator into a selective resonator will
amplify stimulated processes and lead to monochromatic coherent THz
generation.

The spin-polarized current tends to sustain nonequilibrium spin
distribution in the junction. Under such conditions, the stimulated
radiative transitions add power to the spontaneous ones. This leads to
enhancement of the quantum efficiency $\theta$.

The stimulated transitions influences the dependence of the radiation
power on the current. It is seen from Fig.~\ref{fig3} that the power $W$
depends almost quadratically on the current $I$. It should have in mind
that 1) the current polarization is near to 100\% ($P\approx1$), and 2)
the thermal contribution to power $W$ is rather low in our conditions (it
was shown in~\cite{Gulyaev5}, that the power follows the current
variations without inertia). In such a situation, the quadratical
dependence on the current reflects coherent action of radiating spins,
when the wave amplitudes are summed rather than powers.

\section{Conclusions}\label{section6}
In summary, we have evaluated the energy efficiency of the spin injection
oscillator based on a ferromagnetic rod--film structure. It has been shown
that
\begin{itemize}
  \item The material choice influences substantially the efficiency. The
  magnetite films give higher efficiency than Permalloy ones.
  \item An additional element, namely, magnetic flux concentrator,  has
  been proposed to improve efficiency.
  \item The quantum efficiency has been evaluated from the measurements,
  that appeared to be more than unity for magnetite films. This fact is
  interpreted as a manifestation of the stimulated radiative transitions.
  \item The stimulated transitions influence also the form of the current
  dependence of the power, which is close to quadratical one for the
  junction with magnetite film. Under low thermal contribution and 100\%
  spin polarization, the quadratical dependence may be indication to
  coherent action of radiating spins with summation of the wave amplitudes
\end{itemize}

The work was supported by the Russian Foundation for Basic Research,
Grants Nos. 10-02-00030-a and 10-07-00160-a.

\end{document}